\documentclass[final,3p,times,twocolumn]{elsarticle}
\newcommand{\W}{7.5cm}

 \usepackage{graphicx}

\usepackage{amssymb}

\begin{document}
\begin{frontmatter}
\title{Miscible transfer of solute in different types of rough fractures: from random
to multiscale fracture walls heights.}
 \author[label1]{Auradou H.}
\author[label1,label2]{Boschan A.}
\author[label2]{Chertcoff R.}
\author[label1,label2]{D'Angelo M.V.}
\author[label1]{Hulin J.P.}
\author[label2]{Ippolito I.}
 \address[label1]{Univ Pierre et Marie Curie-Paris6, Univ Paris-Sud, CNRS, F-91405.
   Lab FAST, Bat 502, Campus Univ, Orsay, F-91405, France.}
 \address[label2]{Grupo de
Medios Porosos, Departamento de F\'\i sica, Facultad de Ingenier\'\i a,
Universidad de Buenos Aires, Paseo Col\'on 850, 1063 Buenos-Aires, Argentina. }
\begin{abstract}
Miscible tracer dispersion measurements  in
transparent model fractures with different types of wall roughness are reported.
The nature (Fickian or not) of dispersion is determined by
studying variations of  the mixing front as a function of the
traveled distance but also as a function of the lateral scale over
which the tracer concentration is averaged. The dominant convective
dispersion mechanisms  (velocity profile in the gap,
velocity variations in the fracture plane) are established by comparing
measurements using Newtonian
and shear thinning fluids. For small monodisperse
rugosities, front spreading is diffusive with a dominant geometrical
dispersion (dispersion coefficient $D \propto Pe$) at low P\'eclet
numbers $Pe$; at higher $Pe$ values one has either $D \propto Pe^2$
({\it i.e.} Taylor dispersion) for obstacles of height smaller
 than the gap or $D \propto Pe^{1.35}$ for obstacles bridging the gap.
For a self affine multiscale
roughness like in actual rocks and a relative shear displacement $\vec{\delta}$
of complementary walls, the aperture field is channelized in the direction
perpendicular to $\delta$. For a mean velocity $\vec{U}$ parallel to the channels, 
the global front geometry
reflects the velocity contrast between them and is predicted from the aperture field.
For $\vec{U}$ perpendicular to the channels, global front spreading is much reduced.
Local spreading of the front thickness remains
mostly controlled by Taylor dispersion except in the case of a very 
strong channelization parallel to $\vec U$.
\end{abstract}
\begin{keyword}
fractures, roughness, dispersion, multiscale, self-affine, shear-thinning
\end{keyword}
\end{frontmatter}
\section{Introduction} \label{introduction}
The geothermal reservoir of
Soultz-sous-For\^ets, like most geological systems, contains
structures of various size along which flow occurs: three main types
of structures were identified: individual
fractures, fracture clusters and major faults~\cite{Genter97}.
In order to understand these flow systems and help with managerial 
decisions, large scale numerical models incorporating such
heterogeneities have been developed.  Yet, when the transport of solutes 
is  involved, the choice of a dispersion
law (possibly scale dependent) valid at the scale of an individual fracture
remains an open issue~\cite{nas}.\\
At this scale, tracer dispersion results from
 the combined action of the complex
velocity field (varying  both in the gap of the fracture and in its
plane) and of mixing by molecular diffusion. The latter allows the
tracers to move from one streamline to another and homogenizes the
spatial distribution of the tracers. In the
classical approach, tracer particles are assumed to perform a random
walk superimposed over a drift velocity. The latter is the average of the fluid velocity
over an appropriate
volume (the representative elementary volume or REV) while smaller
scale variations induce tracer spreading. At the REV scale, the
average $\overline C(x,t)$ of the tracer concentration over a
section of the medium normal to the mean displacement satisfies the
convection-diffusion equation~\cite{Bear}:
\begin{equation}\label{eq:condiff}
\frac{\partial {\overline C(x,t)}}{\partial t} = U\frac{\partial
{\overline C(x,t)}}{\partial x} +D \frac{\partial^2{\overline
C(x,t)}}{{\partial x}^2}
\end{equation}
where $D$ is the longitudinal dispersion coefficient, $\vec{U}$ the mean
velocity of the fluid (parallel to $x$). The value of $D$ (or equivalently
of the dispersivity $l_d = D/U$)  is independent of both time and the
travelled distance: it is
determined by the combined contributions of molecular diffusion and
advection. The relative order of magnitude of these two effects is
characterized by  the  P{\'e}clet number : $Pe = Ua/D_m$
($D_m$ is the molecular diffusion coefficient;  
$a$ is a characteristic length of the medium (here
the mean fracture aperture).\\
Recent experimental studies of breakthrough curves of solutes in
natural fractures~\cite{Neretnieks82,Park97,Keller2000,Lee2003}
measured dispersion coefficients increasing linearly with the mean
flow $U$ (or with $Pe$). Moreover, the value of the dispersivity
$l_d \ =\ D/U$ observed agreed with the predictions  of a perturbation
analysis~\cite{Gelhar1986}. These results suggested that dispersion
is controlled (like in 3D porous media \cite{Bear}) by spreading due
to velocity variations associated to the geometry of the void
structure. The latter determines the correlation length of the
velocity field, leading to the so called geometrical dispersion
regime. However, flow in fractures is known to be frequently
concentrated in long channels of high hydraulic
conductance~\cite{nas,brown1998,Tsang89}. The velocity remains then
correlated over distances which may be too large for establishing a
Fickian dispersion regime. These previous experiments were all
performed for a fixed path length: however, in order to test the validity of
the Fickian description one must measure
the variation of the width of the mixing front with time $t$ and check whether 
it increases, as expected,  as $t^{1/2}$).\\
  Another key factor is dispersion resulting from the flow
 profile in the gap of the fracture: the  variation of the velocity  between
  the walls (where it cancels out) and the middle of the gap (where it
has a maximal value) stretches the solute front. This creates a
concentration gradient across the gap which is balanced by transverse molecular
diffusion. The decorrelation of the velocity of the solute is then
determined by the characteristic time for the diffusion of solute
particles  across the gap. This differs from the geometrical regime
in which the decorrelation is determined by the geometrical
structure of the fracture.Then, the longitudinal dispersion
coefficient scales like $D\sim Pe^2$ in this so-called Taylor
dispersion (instead of $D\sim Pe$ for geometrical dispersion).\\
In fractures, both dispersion regimes are
expected to coexist (see
refs.~\cite{Ippolito1993,Roux1998,Detwiler2000}): at low
P{\'e}clet numbers (but large enough to neglect pure molecular
diffusion), dispersion is controlled by the disordered geometry,
while, at higher ones, Taylor dispersion becomes the leading
dispersion mechanism. Yet, the critical P{\'e}clet number
characterizing the transition still has to be determined. Also,
the robustness of this model when contact points between the
fracture walls are present must be tested.\\
We discuss in this paper dispersion experiments dealing with these
issues and carried out in transparent fractures with various
 degrees of heterogeneities. The geometries of the void space and the
 roughness of the walls of these models are described in Sec.~\ref{models}.
 They range from a random wall roughness with a correlation length
of the order of the aperture to a multiscale rough wall geometry
similar to that observed in the field~\cite{Sausse2002}; this latter case often leads
 to a strong flow channelization~\cite{Tsang89}.
In the present models a relative shear displacement $\vec{\delta}$ of complementary
 matching rough walls is introduced: high aperture
channels oriented normal to the displacement and spanning over the
fracture are then created leading to an anisotrope aperture
field~\cite{GentierLAR97,Auradou05}. This phenomenon increases with
the magnitude of $\vec{\delta}$ and becomes noticeable as soon as
$\delta$ is of the order of the mean aperture~\cite{matsuki2006}.
The influence of the contact area between the fracture walls was
also investigated by performing flow experiments in a transparent
model fracture with an array of contact points.\\
 In order to address these various issues, dispersion has been studied as a
 function of :\\
 \indent $\bullet$ the distance traveled by the tracer.\\
 \indent $\bullet$ the lateral scale of observation in the
 fracture plane over which the concentration is averaged.
 This scale ranges from a (meso)microscopic scale ({\it i.e.} the
typical fracture aperture) up to the fracture width.\\
\indent $\bullet$ the fluid rheology in order to determine, without
ambiguity, the main mechanisms controlling the dispersion: {\it{i.e.}}
velocity profile in the fracture gap or velocity fluctuations in the fracture plane.\\
This contrasts with previous measurements realized at
the outlet of the samples and in which the  development of
the mixing region and  its spatial structure
cannot be investigated.
\section{Experimental setup and procedure}
\subsection{Experimental models and injection set-up}
\label{models}
$\bullet$ Model $1$: this model (see ref.~\cite{Boschan2008} for details)
has two transparent surfaces of size $350 \times
120\,\mathrm{mm}$ without contact points. The upper one is a flat glass plate
 and the lower
one is a rough photopolymer plate. The wall roughness corresponds to
randomly distributed cylindrical obstacles of diameter $d_o =
1.4\,\mathrm{mm}$ and height $0.35\,\mathrm{mm}$ protruding out of
the plane surface. The minimum aperture $a_m$ of the model is
 the distance between the top of the obstacles and the flat
glass plate with $a_m = 0.37\pm 0.02 \,\mathrm{mm}$; the maximum and
mean values are respectively $a_M
= 0.72 \pm 0.02 \,\mathrm{mm}$ and  $\bar{a}=0.65\pm 0.02\
mm$.\\
\indent $\bullet$ Model $2$: this model uses a periodic square array of obstacles of similar size as in
model $1$ but of rectangular and variable cross section and with their top
 in contact with the top plate. Flow takes then place in a two
dimensional network of channels of random aperture (see
ref.~\cite{Dangelo2007} for a full description). The model
contains $150 \times140$ channels (real size $150 \times 140\
\mathrm{mm}$) with an individual length equal to $l \ = \ 0.67\
\mathrm{mm}$ and a depth $a_M = 0.5~\mathrm{mm}$; their average width is
$\bar{w}=0.33\ \mathrm{mm}$ and its standard deviation $\sigma(w)\
=\ 0.11\ \mathrm{mm}$. Following the definition of Bruderer and
Bernabe~\cite{bruderer01}, the degree of heterogeneity of the
network can be characterized by the normalized standard deviation
$\sigma(w)/\bar{w}$. In the present work : $\sigma(w)/\bar{w}
\simeq 0.3$.\\
\indent $\bullet$  model $3$: Models $3$, $4$ and $5$ have complementary self-affine
walls of size $350 \times 90\ \mathrm{mm}$, reproducing the roughness of natural fractures  (see
ref.~\cite{Boschan2007}). In model $3$, a relative shear
displacement $\delta = 0.75\ \mathrm{mm}$ parallel to the
direction of the flow is applied between the walls. the mean of the fracture aperture is
$\bar{a}=0.75\ \mathrm{mm}$ and its standard deviation is
$\sigma_a=0.11\ \mathrm{mm}$. This shear configuration is referred to
as $\vec{\delta} \parallel \vec{U}$.\\
\indent $\bullet$  model $4$:
In order to analyze the influence of the direction of the shear displacement,
 the direction of the shear for model $4$ is now perpendicular to the direction of
the flow (the corresponding standard deviation of the aperture is
$\sigma_a=0.15\ \mathrm{mm}$). This configuration (and that of model
$5$) is referred to as $\vec{\delta} \perp \vec{U}$. All other
characteristics  (wall size, mean  aperture, map of the roughness of
each wall, amplitude $\delta  = 0.75\ \mathrm{mm}$
of the shear displacement) are identical to those of model $3$.\\
All models are transparent and placed vertically with their open
sides horizontal. The upper side is fitted with a leak tight adapter
allowing one to suck the fluids at a constant flow rate. The lower
open side can be dipped into a bath containing the liquid. When the
pump is switched off, the bath can be lowered before changing
the fluid inside it. This allows one to obtain a flat
initial front between the fluids (See Figure $1$ in
ref.~\cite{Boschan2007}).\\
The models are illuminated from the back by a light panel and images
are acquired using a high resolution camera. The pixel size is
around $0.2\ mm$, {\it i.e.} lower than the typical fracture
aperture. About $100$ images of the distribution of the light
intensity $I(x,y,t)$ transmitted through the fracture are recorded
at constant intervals during the fluid displacement using a digital
camera with a high dynamic range. Reference images with the fracture
saturated with the clear and dyed fluids (dye concentration $c_0$)
are also recorded before the experiments and after the full
saturation by the displacing fluid. A calibration curve obtained
independently through separate measurements is then used to map the
local relative dye concentration $0 \le c(x,y,t)/c_0 \le 1$ (in the
following, $c_0$ is omitted and $c(x,y,t)$ refers directly to the
normalized dye concentration). The two fluids are of equal density: 
this is verified by performing twice the experiments at each flow rate value with the dyed fluid either
displacing or displaced by the clear fluid. Comparing the results allows one to 
 to detect possible instabilities induced by residual
density differences (the corresponding experiments are discarded). The two fluids
are, of course,  miscible and have the same viscosity.
\subsection{Fluids preparation and characterization}
\label{fluid}
The solutions used in the present work are either a Newtonian
water-glycerol mixture or shear-thinning water-polymer
(scleroglucan) solutions with a $500$ or $1000\, \mathrm{ppm}$
polymer concentration. In all cases, the injected and displaced
fluids have identical rheological properties.
The Newtonian solution
contains $10\%$ in weight of glycerol and has a viscosity equal to
$1.3 \times 10^{-3}\, \mathrm{Pa.s}$ at $20^\circ C$.
\begin{table} \label{tab1}
\begin{tabular}{lccccc}
$Fluids$ & $n$ & $\dot{\gamma_0}$ & ${\mu}_0$\\
           &     & $s^{-1}$       & $mPa.s$ \\
$W-Glycerol$&$1$&$-$&$10$\\
$500\ ppm$ & $0.38 \pm 0.04$ & $0.077 \pm 0.018$ & $410 \pm 33$\\
$1000\ ppm$ & $0.26 \pm 0.02$ & $0.026 \pm 0.004 $ & $ 4500 \pm 340$\\
\end{tabular}
\caption{Rheological parameters and P\'eclet numbers
for the $500$ and $1000\ ppm$ 
scleroglucan solutions solutions used in the present work. $W-Glycerol$
refers to the water glycerol mixture.}
\end{table}
The preparation and characteristics of the shear-thinning
solutions are the same as reported in ref.~\cite{Boschan2007}. The
variation of the viscosity $\mu$  with the
 shear rate $\dot{\gamma}$ is well fitted by the  Carreau function:
\begin{equation}\label{powervisc}
\mu = \frac{1}{(1+
(\frac{\dot{\gamma}}{\dot{\gamma_0}})^2)^{\frac{1-n}{2}}} (\mu_0 -
\mu_{\infty}) + \mu_{\infty}.
\end{equation}
The values of the rheological parameters characterizing the fluids
are listed in Table~1. For the non Newtonian fluids and
at low shear rates $\dot{\gamma} \lesssim \dot{\gamma_0}$, the
viscosity is constant like for a Newtonian fluid with $\mu \simeq
\mu_0$ (Newtonian plateau regime). At higher shear rates
$\dot{\gamma} \gtrsim \dot{\gamma_0}$, the 
viscosity follows a power law: $\mu \propto
{\dot{\gamma}}^{(n-1)}$.  Practically, $\mu_{\infty}$ is taken
equal to $1\, \mathrm{mPa.s}$, {\it i.e.} the viscosity of
water (the solvent): this limiting value would indeed only be
reached at  shear rates above the experimental range. \\
The two main dispersion mechanisms {\it{i.e.}} Taylor dispersion ($D\sim Pe^2$) 
and geometrical dispersion ($D\sim Pe$) are affected in opposite directions 
when a Newtonian fluid is
replaced by a shear thinning solution. More precisely, the
velocity contrasts between different flow paths are enhanced for a
shear thinning fluid, resulting in an increase of the geometrical
dispersion (without modifying the scaling law $D\sim Pe$). By
contrast, the velocity profiles in the gap become flatter: this reduces
therefore Taylor dispersion, but still with $D \propto Pe^2$. Varying the fluid
rheology modifies  the relative influence of the
two main dispersion mechanisms in opposite ways:
 the dominant one can therefore be identified
unambiguously for each fracture
geometry and flow rate.
\section{Experimental results}
\subsection{Fracture model $1$} \label{sec:mod1}
\begin{figure}
\includegraphics[width=\W]{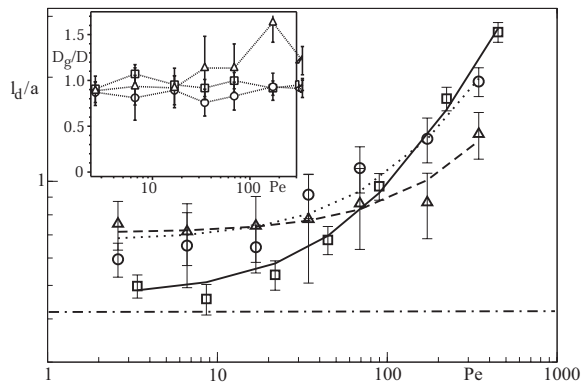}
\caption{Variation of the experimental dispersivity $l_d$ as a
function of the P\'eclet number in model $1$. ($\square$): water-glycerol
solution; ($\triangle$):  $1000$\ ppm, ($\circ$): $500$\ ppm
polymer solutions. Solid, dotted and dashed lines: fit of the
respective data with
 Eq.~(\ref{eq:model}).}\label{fig:figure1}
\end{figure}

In this model,  flow takes place  in the free space between a flat
plate and a second one with protuberant obstacles. The latter
perturbs the flow velocity field: the local mean fluid velocity
(averaged over the gap) is greater between the obstacles, where the
aperture is largest than at their top, where it is minimal. These
mean velocity variations in the fracture plane result in geometrical
tracer spreading. As for the velocity profile in the fracture gap,
it induces Taylor like dispersion. The variation of the dispersivity
$l_d = D/U$ as a function of  $Pe$ confirms that it is the sum of
the two  contributions discussed above with:
\begin{equation}\label{eq:model}
\frac{l_d}{a}=\alpha_G + \alpha_T Pe,
\end{equation}
where  $\alpha_T Pe$  corresponds to Taylor dispersion and
$\alpha_G$ to geometrical dispersion. For a fracture with two flat
parallel plates and  a Newtonian fluid, one has: $\alpha_G = 0$
and $\alpha_T = 1/210$; also, one has $\alpha_G \neq 0$  only for
fractures with rough walls. Moreover, if the correlation length of
the velocity field is small compared to the fracture size and if
the ratio $\epsilon$ of the amplitude of the velocity
fluctuations  to the mean velocity $U$ is small, then the
perturbation theory predicts that $\alpha_G \propto \epsilon^2$ (a
complete expression of $\alpha_G$ is given by
Eq.~($3$) of ref.~\cite{Boschan2008}).\\
Experimental dispersivity variations as a function of $Pe$ are
plotted in Fig.~\ref{fig:figure1} for the three fluids. These data
sets are well adjusted (see lines in Fig.~\ref{fig:figure1}) by
functions of the type shown in Eq.~(\ref{eq:model}): the
dispersivity increases at first slowly with $Pe$ above $Pe \simeq
20$ from a nearly constant plateau value before displaying a linear
variation at higher velocities. The plateau value corresponds to
$\alpha_G$ in Eq.~(\ref{eq:model}) and increases with the polymer
concentration. It can be shown that the amplitude of the velocity
fluctuations is larger for  shear thinning fluids: for a power law 
dependence of the viscosity on the shear rate
($\mu \propto \dot{\gamma}^{(n-1)}$), the parameter $\epsilon$
would increase theoretically by a factor $(1+1/n)/2$ compared
 to a Newtonian fluid. The
velocity fluctuations (and, as a result, the dispersivity) increase
therefore when $n$ decreases {\it i.e} when the shear thinning
character of the fluids is stronger. Unlike $\alpha_G$, the parameter $\alpha_T$ for
shear-thinning fluids is  lower than the Newtonian value  $1/210$
 (see \cite{Boschan2008}).\\
The values displayed in Fig.~\ref{fig:figure1} were  obtained by
fitting the local concentration variation on each individual pixel
by solutions of Eq.~(\ref{eq:condiff}). A similar analysis was
performed on the average of the local concentrations over the
fracture width. The results are displayed by empty symbols in
Fig.~\ref{fig:figure1}: they almost fall on the filled symbols
demonstrating the lack of large scale heterogeneities in this model
fracture.
\subsection{Fracture model $2$}
\begin{figure}
\includegraphics[width=\W]{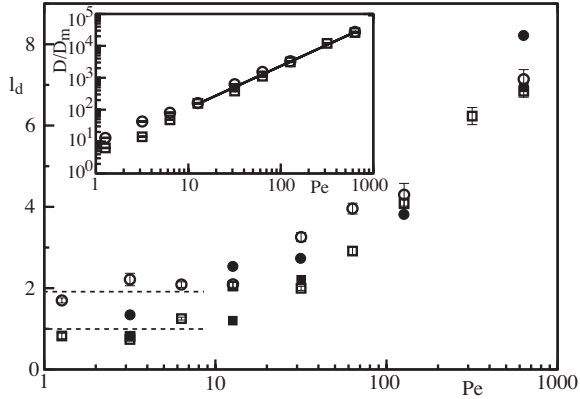}
\caption{Variation of the dispersivity $l_d$ ($mm$) with the
P{\'e}clet number for experiments with water-polymer solutions :
($\square$),($\blacksquare$) : $500 \ ppm$ concentration  -
($\circ$), ($\bullet$) $1000\ ppm$. Open (resp. filled) symbols :
averaging interval : $35$ (resp. $0.4$) mesh sizes. Dashed lines :
Mean dispersivity values for the geometrical dispersion regime.
Dotted line: power law fit of the variation for $Pe>10$ (exponent
$0.35 \pm 0.03$).}\label{fig:figure2}
\end{figure}

In this model, the obstacles extend over the full gap
height and mimic gouge particles created by the failure of the
rock and  evenly distributed in the
fracture (see Sec.~\ref{models}). The model appears then as a plane
array of channels of random width: it can be considered as a $2D$
porous medium in which the pores correspond to the junctions between
the channels.  We show now that mixing at these pore junctions
has a crucial influence on dispersion.\\
Fig.~\ref{fig:figure2} displays variations of the dispersivity $l_d$
with the P\'eclet number deduced from time variations of the
local concentration at the pore scale (filled symbols) and of its average 
over the fracture width (open symbols); it is seen that  the values of $l_d$
obtained in both cases are similar so that, in the following, only global 
measurements will be discussed.\\
For $Pe<10$, $l_d = D/U$ is nearly constant ({\it i.e.} $D \propto Pe$), suggesting
dominantly geometrical dispersion. As discussed 
in Sec.~\ref{sec:mod1}, the value of $l_d$ in this regime should depend strongly on
the rheology of the solution: more precisely, it should increase
with  the polymer concentration as indeed observed
here (like for model $1$).\\
For $Pe > 10$, a second dispersion regime is observed, in which 
$l_d$ increases with $Pe$. Furthermore,
the linear trend observed in a log-log coordinate shows that $l_d$
follows a power law of $Pe$ (more precisely, $l_d \propto Pe^{0.35}$
for $Pe > 10$). This result is in agreement with numerical
simulations  by Bruderer and Bernabe~\cite{bruderer01} and differs
from the Taylor dispersion regime $l_d \propto Pe$ observed in
model $1$ at high $Pe$ values.\\
This difference is explained by the influence of the pore junctions.
At low  flow velocities (typ.  $Pe < 10$),  tracer particles
can explore effectively the local flow field by molecular diffusion
during their transit time through a given  junction: this distributes evenly the tracer
concentration inside it which represents
a perfect mixing condition. Then, the tracer concentration is
equal in all outgoing paths  and the probability to follow 
one of them is proportional to the corresponding flow rate~\cite{adler99,park01}. 
Therefore, in this regime, dispersion is
controlled by the disordered geometry of the array of channels.\\
 At higher $Pe$ values (typ. $Pe > 10$), mixing at the junctions is
 no more perfect and the tracer concentration in slower channels
(like those transverse to the mean flow) is lower compared to the
perfect mixing situation. The dispersion characteristic becomes more
similar to the case of capillary tubes (representing the fast flow
channels) oriented along the flow direction. In this case, one would
observe Taylor dispersion with $l_d \propto Pe$ (or $D\propto Pe^2$) but  the
influence of flow redistribution at the junctions is quite large: this
leads to a variation of $l_d$ as $Pe^{0.35}$ intermediate
between those observed in the geometrical
and Taylor regimes.
\subsection{Fracture model $3$}
\begin{figure}
\includegraphics[width=\W]{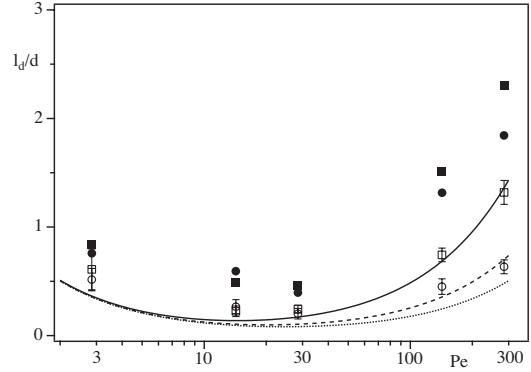}
\caption{Variation of the normalized dispersivity $l_d/d$ as a
function of $Pe$ for model $3$ and two different
polymer concentrations.
($\bullet, \blacksquare$): global dispersivities determined from concentrations
averaged over the
fracture width. ($\circ,\square$): local dispersivities  determined from 
concentration variations
on individual pixels. Lines: Taylor dispersion for
plane parallel walls with the same mean gap as for model $3$.
($\circ,\bullet$), dotted line: $1000 \mathrm{ppm}$ polymer solution;
($\square,\blacksquare$), dashed line: $500 \mathrm{ppm}$
polymer solution; continuous line: Newtonian solution. Insert: variation of the
ratio of the local and global dispersivities as a function of the P\'eclet 
number. ($\square$): 
$500$ \ ppm solution. ($\circ$): $1000$ \ ppm solution.} \label{fig:figure3}
\end{figure}

Like in model $1$, the walls of this fracture do not have any
contact point but, in contrast with it, the rugosities of the wall
have been selected to reproduce  the multi-scale roughness of most
natural fractures (see  Sec.~\ref{models}).\\
Such fractures are known to display high aperture channels
perpendicular to the relative shear displacement $\vec{\delta}$ of the walls;
they  are characterized by an anisotropic permeability field with
a larger permeability  in the direction parallel to the channels.
While most studies of these systems have dealt with their
permeability, little  is known about the influence of such a structure
 on tracer dispersion.\\
In model $3$,  flow is parallel to $\vec{\delta}$ ({\it i.e.} normal
to the channels): in this case ($\vec{\delta} \parallel \vec{U}$),
both the local and global concentration variation curves are  well
adjusted by the solution of the convection-dispersion equation~(\ref{eq:condiff}).
 Moreover, dispersivity values determined
from these curves become constant after a long enough
path inside the fracture. Like
for models $1$ and $2$, the dispersion process is therefore Fickian.
Fig.~\ref{fig:figure3} displays  variations of both the local and global
dispersivities with $Pe$ for the two polymer solutions. 
Theoretical Taylor
dispersivities for a fracture of same mean aperture with
plane smooth walls and for the different fluid rheologies are also
plotted in Fig.~\ref{fig:figure3} as dashed and dotted lines (differences
between these curves
reflect the effect of the velocity profile in the gap).\\
For $Pe > 12$, the  local dispersivity  increases with $Pe$
in qualitative agreement
with theoretical expectations ($l_d=D/U\sim Pe$) and is also lower for  the
strongly shear-thinning $1000~\mathrm{ppm}$ solution (open symbols). 
For both solutions  $l_d$ is larger than predicted, particularly for 
the $500$ \ ppm solution for which it is close to the Newtonian value.
This may be due to the vicinity of the ``plateau'' domain of the rheological curve
 in which the solution behaves like a Newtonian fluid at low shear rates. 
 For $Pe \sim 12$, in which both solutions should be in this ``plateau'' regime,
  the dispersivities  are, as expected, the same for the two solutions but
  still slightly higher than the theoretical  value. At $Pe < 10$, $l_d$ rises again
  due to the influence of longitudinal molecular diffusion and its value is 
  also the same for the two solutions (the ($\circ$) and ($\square$) symbols coincide).\\
These value of the local dispersivity are compared in Fig.~\ref{fig:figure3}
to the global dispersivities  determined from time variations  of
the  concentration averaged over the fracture width (filled symbols):
as seen in Fig.~\ref{fig:figure3} and its inset, the local dispersivities 
are significantly smaller
 (at a same P\'eclet number and for a same solution).
 The front contours ($c = 0.5$) displayed in
Figs.~\ref{fig:figure4}a and b for model $3$ reveal  fine structures of the mixing
front: they reflect  fluctuations of the velocity induced by the fracture wall
roughness. Their magnitude is large enough to account for the additional increase of the
global dispersivity with respect to pure Taylor dispersion (compared to local dispersion)
 but not enough to allow for the observation of a geometrical dispersion regime.\\
To conclude,  in model $3$ with $\vec{\delta} \parallel \vec{U}$,
dispersion
 is mostly controlled by the
Taylor dispersivity component due to the velocity profile between the walls
as soon as $Pe \gtrsim 12$; there is however an amplification of the dispersion due
 to the fracture roughness.
\subsection{Fracture model $4$}
In model $3$, the mean
flow was perpendicular to the channels or to the ridges induced by the
shear displacement: the correlation length of the velocity is then
determined by  the typical width of these structures.
Model $4$ has the same size as model $3$, a same mean aperture and
complementary
rough walls with a self-affine geometry exactly identical to that
used for model $3$. However, the shear $\vec{\delta}$
is, this time, perpendicular to the mean flow $\vec{U}$.
In this configuration ($\vec{\delta} \perp \vec{U}$), $\vec{U}$ is parallel to the channels 
and ridges created
by the shear: the correlation length of the flow velocity is then
 determined by the length of the channels which
is much larger than their width.\\
\begin{figure}
\center {\includegraphics[width=\W]{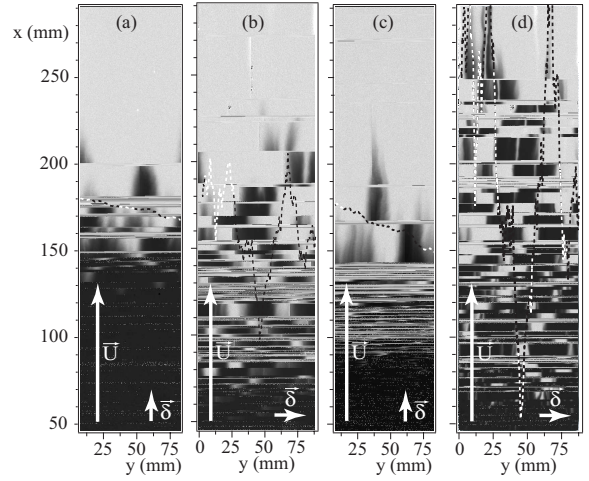}} \caption{Experimental isoconcentration fronts
 ($c = 0.5$) as a function of the normalized distance $x/\overline{x(t)}$ ($\bar{x}$ = mean front distance )
for different ratios $\alpha$ of the injected volume to the pore volume (a)-(b): fracture
model $3$ ($\vec{\delta} \parallel \vec{U}$);
 (c)-(d): fracture model $4$ ($\vec{\delta} \perp \vec{U}$). Mean velocities: (a)-(c):
 $U = 0.0125 \ mm/s$, $Pe = 14$; (b)-(d): $U = 0.25 \ mm/s$, $Pe = 285$. 
 (dots: $\alpha= 0.85$, dash-dots: $\alpha = 0.65$, dash-dot-dots :  $\alpha = 0.5$, dashes:
  $\alpha = 0.15$). Continuous line: theoretical variation from Eq.~\ref{eq:xfront}. 
All experiments have been realized with identical
$1000\, \mathrm{ppm}$ water-polymer solutions.}\label{fig:figure4}
\end{figure}
The dispersion characteristics are then very different as can be seen
by comparing  isoconcentration fronts obtained
 for model $4$ (Figs.~\ref{fig:figure4}c-d) and model
$3$ (Figs.~\ref{fig:figure4}a-b) at different times and
 in identical experimental conditions.
More precisely, large fingers and troughs are observed for model $4$
while none appears for model $3$. Also, the amplitude of these 
features parallel to $\vec{U}$ is larger at the higher velocity  for which
the solution has a shear-thinning behaviour (Fig.~\ref{fig:figure4}d)
 than at the lower velocity
at which it behaves like a Newtonian fluid (Fig.~\ref{fig:figure4}d).
Another important feature is the good collapse of the large features 
of the front observed at different times when normalized by the mean
distance: this shows that the size of these features parallel to the flow
increases linearly with time.  Such a collapse is not apparent in model
$3$, except near the sides of the model where they likely reflect wall effects.\\
These results show that front spreading
is purely convective and that the total width $\Delta x$ of the front
parallel to $\vec{U}$ ({\it i.e.} the distance between the tips of
the fingers and the bottom of the troughs) increases linearly with
distance as $x\ \Delta U/U$ ($\Delta U/U$ = typical large scale
velocity contrast between the different channels  created by the shear). \\
In order to predict these contrasts,
we modelled the fracture aperture field as a set of independent parallel
channels of aperture
$a(y) = <a(x,y)>_x$~(\cite{Auradou06,Auradou08}). A particle
starting at a transverse distance $y$ at the inlet is assumed to
move at a velocity proportional to $a(y)^{(n+1)/n}$;
 The theoretical profile $x_f(y,t)$ of the front at a time $t$ is then:
\begin{equation}\label{eq:xfront}
x_f(y,t)=\frac{\overline{x(t)} \
a(y)^{(n+1)/n}}{<a(y)^{(n+1)/n}>_y},
\end{equation}
where $\overline{x(t)} = <x_f(y,t)>_y$ and  $<a(y)^{(n+1)/n}>_y$ are
averages  over $y$ of the local aperture $a(x,y)$. Normalized
profiles $x_f(y,t)/\overline{x(t)}$ computed using
Eq.~(\ref{eq:xfront}) and  the actual aperture fields  are plotted  in
Figs.~\ref{fig:figure4}a to d as continuous lines. The exponent $n$ has 
been taken equal to $1$
at the lowest velocity for which $\dot{\gamma} \sim \dot{\gamma_0}$
(Figs.~\ref{fig:figure4}a-c) and to $0.26$ at the highest one for which
$\dot{\gamma} > \dot{\gamma_0}$  (Figs.~\ref{fig:figure4}b-d) (as mentioned
in Sec.~\ref{fluid}, $\dot{\gamma_0}$ is the shear-rate value corresponding to 
the crossover from the Newtonian to the shear-thinning behaviour of the fluid).\\
Eq.~(\ref{eq:xfront}) clearly predicts well the location and shape of the
large ``fingers'' and ``troughs'' at both velocities for $\vec{\delta}
\perp \vec{U}$.
In contrast,  the theoretical curves does not reproduce
the front geometries in model $3$ ($\vec{\delta} \parallel \vec{U}$) except
for the small global slope.\\
This confirms that, for $\vec{\delta} \perp \vec{U}$ (model $4$),
 the large scale
 features of solute transport are determined by the velocity contrasts between
the channels created by the shear. 
The curves of Figs.~\ref{fig:figure4}c-d also reproduce well the difference
between the sizes of the fingers at the two velocities investigated. This
confirms that  the difference between these sizes may be accounted
for by the different rheological behavior of the fluid : the velocity contrasts
 (and, therefore, the
size) are amplified for $Pe = 285$ (shear-thinning power law domain)
compared to the vicinity of the Newtonian constant viscosity regime
($Pe = 14$).\\
For model $3$,  the hypothesis of the model are not satisfied and
it does not reproduce the front geometry: however,
 the features of the front  are generally
visible at similar  transverse distances $y$ in
Figs.~\ref{fig:figure4}a and \ref{fig:figure4}b (at a given time). They reflect likely also
in this case a convective spreading of the front  due to velocity contrasts between
 the  flow paths: however, there is no simple relation of the front geometry
 to the aperture field, in contrast with model $4$.\\
The local dispersivity $l_d(x,y)$ has also  been determined for model $4$ 
from the variations of the concentration on single
pixels: its values are overall larger and their   distribution  is much broader than for 
model $3$.\\
 The same measurements have been performed~\cite{Boschan2007} on 
a model fracture with a similar wall geometry but with a smaller 
amplitude $\delta = 0.33\ \mathrm{mm}$   of the shear (still with $\vec{\delta} \perp \vec{U}$).
 In this case,  the values
of the local dispersivity are very close to those  predicted from
 Taylor dispersion. Thogether with a smaller amplitude of the large scale fingers, this 
 reflects   a weaker disorder of the flow field.                                                   
\section{Discussion}
The experiments reported in this  paper for several model fractures demonstrate the
key influence of  wall roughness geometries on the dispersion processes and their dependence on $Pe$
and on the fluid rheology. One can group the results in two sets: \\
\indent $\bullet$ models $1$ and $2$: both models correspond to obstacles
with a single characteristic size.
The height of the obstacles is smaller than the aperture for model $1$
and equal to it in model $2$: this models  the case of gouge (or proppant)
particles bridging the gap.
In both cases the variation with distance and time
of the tracer concentration satisfies  the convection
dispersion equation~(\ref{eq:condiff}); the values of $D$ are independent of
the fraction of the width of the model over which the concentration is averaged
and also of the distance from the inlet.\\
At low  P{\'e}clet numbers, one has, in both cases, $D \propto U$  corresponding to geometrical dispersion
due to the disorder of the velocity; in this regime, $l_d = D/U$
increases with the polymer concentration ({\it i.e.} with the shear-thinning character of the fluids)
due to an enhancement of the velocity contrasts.
Moreover, for model $1$, the value of $l_d$ is close to that predicted from a small perturbation theory.
At high P{\'e}clet numbers,
there is, for model $1$  a transition towards
Taylor dispersion with $D\propto Pe^2$.
In model $2$,  $D$ increases at high $Pe$ values as $Pe^{1.35}$:
this exponent agrees with previous numerical simulations~ \cite{bruderer01} and
 should depend on the distribution of the size of the obstacles.
In model $2$, the transition between the different  regimes  is
controlled by mixing at the scale of individual junctions.\\
\indent  $\bullet$ models $3$, $4$ and $5$:
The roughness of the walls of these models has a multiscale self-affine geometry
similar to that of many fractured rocks; the walls of these fractures are complementary
with a relative shear displacement either parallel ($\vec{\delta} \parallel \vec{U}$) for model $3$ or
perpendicular ($\vec{\delta} \perp \vec{U}$) for models $4$ and $5$.
The relative shear produces a channelization perpendicular
to $\vec{\delta}$ of the aperture field: as a result,  dispersion depends strongly on the relative
orientation of $\vec{\delta}$ and $\vec{U}$.\\
For $\vec{\delta} \perp \vec{U}$,  the global spreading of the mixing front is not
dispersive. The global width of the front  parallel to $\vec{U}$ increases instead linearly with time
and reflects  directly the velocity contrasts between the channels
created by the shear. The large scale structures of the front can be predicted 
 from the aperture field and their size increases with the shear-thinning character of the fluid. 
 The variation of the local thickness of
the front remains instead dispersive, but  with a magnitude larger than for Taylor dispersion.
For model $3$ ($\vec{\delta} \parallel \vec{U}$), the global spreading of the front is much weaker
that in model $4$ which has the same characteristics but for which $\vec{\delta} \perp \vec{U}$:
 local spreading is controlled
by Taylor dispersion at large $Pe$'s and by molecular diffusion at lower ones.
\section{Conclusion}
The experiments reported here demonstrate that varying the fluid rheology is a powerful diagnostic
 tool  for understanding hydrodynamic tracer dispersion mechanisms  in  rough fractures.\\
For models $1$ and $2$, both the size of the wall rugosities and the correlation length of the
velocity field are small compared to the global size of the fracture: this allows one to reach
a geometrical dispersion regime at low $Pe$ values. At higher $P$e's, other characteristics
of the structure of the void space such as the flow profile in the aperture (model $1$) and
the distribution of the tracer in the pore junctions (model $2$) strongly influence dispersion.
In these models, dispersion may be characterized by a single macroscopic dispersion coefficient:
however the knowledge of the microscopic structure of the fracture aperture field (correlation length, 
pore size...) is necessary to predict its value and dependence on $Pe$.\\
Experiments performed on multiscale fractures (models $3$ and $4$) reproducing the 
roughness of natural fractures have demonstrated the strong influence of channelization
and of its orientation with respect to the mean flow on the transport of tracer.
An important issue is whether, in these cases, transverse exchange of tracer is large enough
so that a diffusive spreading regime might be reached at very large distances.
These results have potentially a strong relevance to the efficiency of the recovery
  of heat through water circulation in geothermal reservoirs. There are also other
possible applications to the prediction of seismic events from water circulation
 in the rock layers under stress.

\end{document}